\documentclass{aastex}
\usepackage{spr-astr-addons}
\usepackage{url}
\usepackage{graphicx}
\usepackage{subfigure}
\usepackage{afterpage}
\usepackage{placeins}

\RequirePackage{color}

\begin{document}

\title{A Systematic Study on Energy Dependence of Quasi-Periodic Oscillation Frequency in GRS 1915+105}
\shorttitle{Energy Dependence of QPO Frequency in GRS 1915+105}
\shortauthors{Yan et al.}

\author{Shu-Ping Yan\altaffilmark{1,2,3,4}}%\email{yanshup@uao.ac.cn}
\author{Jin-Lu Qu\altaffilmark{2,4}}%\email{qujl@ihep.ac.cn}
\author{Guo-Qiang Ding\altaffilmark{1}}%\email{dinggq@uao.ac.cn}
\author{Peng Han \altaffilmark{2,4}}
\author{Li-Ming Song\altaffilmark{2,4}}%\email{songlm@ihep.ac.cn}
\author{Hong-Xing Yin\altaffilmark{5}}
\author{Cheng-Min Zhang\altaffilmark{6}}
\author{Shu Zhang \altaffilmark{2,4}}
\author{Jian-Min Wang\altaffilmark{2,4}}

\altaffiltext{1}{Xinjiang Astronomical Observatory, Chinese Academy of Sciences, 150, Science 1-Street, 
Urumqi, Xinjiang 830011, China; dinggq@uao.ac.cn, yanshup@uao.ac.cn}
\altaffiltext{2}{Key Laboratory of Particle Astrophysics, Chinese Academy of Sciences, 19B Yuquan Road, 
Beijing 100049, China; qujl@ihep.ac.cn, songlm@ihep.ac.cn}
\altaffiltext{3}{Graduate University of Chinese Academy of Sciences, 19A Yuquan road, Beijing 100049, China}
\altaffiltext{4}{Opening Laboratory of Cosmic ray and High Energy Astrophysics, 19A Yuquan road, Beijing 100049, China}
\altaffiltext{5}{School of Space Science and Physics, Shandong University, 264209, Weihai, China}
\altaffiltext{6}{National Astronomical Observatoires, Chinese Academy of Sciences, 100012, Beijing, China}

\begin{abstract} 

Systematically studying all the {\it RXTE}/PCA observations for GRS 1915+105 before November 2010, we have discovered three additional patterns in the relation between Quasi-Periodic Oscillation (QPO) frequency and photon energy, extending 
earlier outcomes reported by \citet{Qu10}. We have confirmed that as QPO frequency increases, the relation evolves from 
the negative correlation to positive one. The newly discovered patterns provide new constraints on the QPO models.

\end{abstract}


\section{Introduction}

GRS 1915+105, discovered by WATCH instrument on board {\emph{GRANAT}} in 1992 \citep{Castro92} and located in our galaxy at an estimated distance of $9\pm3$ kpc \citep{Chapuis04}, is a low-mass X-ray binary containing a spinning accreting black hole \citep{Zhang97} of mass about $14\pm4$ M$_{\odot}$ and a K-M III giant star of mass $0.8\pm0.5$ M$_{\odot}$ as the donor \citep{Harlaftis04, Greiner01a}. The orbital separation and period of this binary are, respectively, about $108\pm4$ 
R$_{\odot}$ and 33.5 days \citep{Greiner01b}. Serving as a famous microquasar, GRS 1915+105 produces superluminal radio jets \citep{Mirabel94, Fender99}. It shows various X-ray light curves and complex timing phenomena. Based on the appearance of light curves and color-color diagrams, the behaviors of GRS 1915+105 can be classified into 12 classes. The variability of the source can be further reduced to transitions between three basic states (A, B, and C) \citep{Belloni00}. Of these 12 classes, class $\chi$ (state) is most commonly observed \citep{Belloni00}. It shows characteristics exclusively of state C, the state which is steady in the X-rays and lies in a rather hard part of the color-color diagram. It is the state when the low-frequency ($\sim0.5-10$ Hz) QPOs (LFQPOs) are most frequently observed \citep[e.g.,][]{Muno99}, providing an idea site for studying LFQPOs.

Much effort has been made for exploring the origins of the LFQPOs of GRS 1915+105. It was found that the QPO frequency was positively correlated with the fluxes of the individual components and spectral total flux \citep[e.g.,][]{Chen97, Markwardt99, Muno99, Trudolyubov99, Reig00, Tomsick01}. \citet{Muno01} confirmed that QPO frequency was tightly correlated with the source flux, and, however, they also found that for some observations the QPO frequency was not correlated with the flux. It was found that the QPO amplitude was inversely correlated with the source flux or QPO frequency \citep[e.g.,][]{Muno99, Reig00, Trudolyubov99}. \citet{Muno99} and \citet{Rodriguez02a} reported that as QPO frequency increased, the temperature of the inner accretion disk increased and the radius of the inner accretion disk decreased. These results indicate that the LFQPO is linked to both the accretion disk and the region where the power law component is produced. However, most of these results are related to models. As a model-independent means, it is meaningful to study the correlations between photon energy and QPO parameters including its amplitude and frequency. Some authors found that the QPO amplitude increased with photon energy and it turned over in high energy bands in some cases \citep[e.g.,][]{Tomsick01, 
Rodriguez02b, Rodriguez04, Zdziarski05}. \citet{Qu10} studied the LFQPOs of GRS 1915+105 in class $\chi$ state \citep{Belloni00} and found that as the centroid frequency of QPO increased the correlation between QPO frequency and photon energy evolved from a negative correlation to a positive one. 

Nevertheless, systematic studies on the energy dependence of the LFQPO frequency in GRS 1915+105 have never been done. In this work, using all the data of {\it RXTE}/PCA of GRS 1915+105 before November 2010, we have investigated the correlation between photon energy and QPO frequency throughout. The data reduction methods are described in \S 2, the results are presented in \S 3, while a simple discussion and the conclusion are given in \S 4.

\section{Observations and Data Reduction}

We analyze all the {\emph{RXTE}} observations of GRS~1915+105 before November, 2010, which are listed in Table \ref{table1}. These observations are belonged to class $\chi$ state and with abundant LFOPOs (0.5--10 Hz) for evaluating the energy dependence of QPO frequency.

\begin{figure}[htbp]
\centerline{\includegraphics[height=8cm,angle=-90]{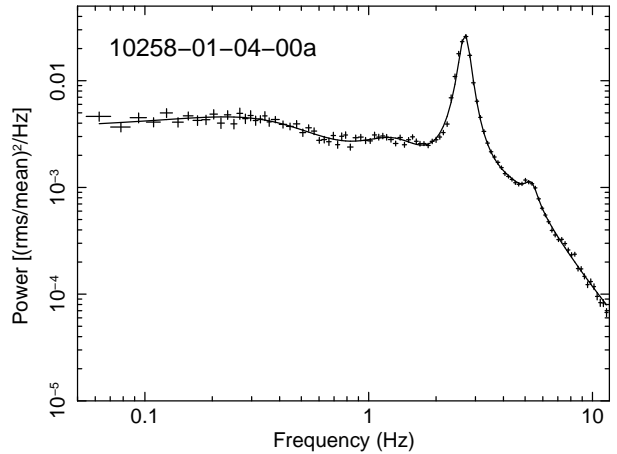}}
\caption{An example of five-Lorentzian model fit of the PDS. The PDS are produced in PCA absolute channel 0-35.}
\label{fig:pds} 
\end{figure}

The light curves are extracted from the binned mode and event mode data of {\it RXTE}/PCA by using the HEASOFT version 6.7 package. Good time intervals are defined as follows: satellite elevation over the Earth limb $>10^{\circ}$ and offset pointing $<0.02^{\circ}$. In order to acquire the details of the correlations between photon energy and QPO frequency 
in broadband with enough confidence, only the generic binned configuration data with energy channel number $\geq 4$ and time resolution $\leq 8$ ms are selected. The light curves are extracted with a time resolution of 8 ms in PCA energy bands defined in Table \ref{table2}. By running POWSPEC version 1.0 with ``normalization~=~-2" option, the power density spectra (PDS) are produced with the normalization of \citet{Miyamoto92}, which gives the periodogram in units of (rms/mean)$^2$/Hz, and corrected for Poisson noise \citep[for details on PDS computation and X-ray PDS normalization practice, see, e.g.,][]{Klis89, Vaughan03}. The PDS are computated on an interval length of 64 s and Logarithmically rebinned by inputting -1.03 to rebin option. Following \citet{Belloni02}, we fit the PDS with a model including several Lorentzians to represent the QPOs, the continuum, and other broad features, respectively. The continuum of PDS can also be fitted with other models, for instance, a power law or a doubly broken power law \citep{Belloni90}. We have compared the quality of our current fits with that of the model including a power law for PDS continuum and several Lorentzians for other timing features and found that the multi-Lorentzian model fits better. A model consisting of a doubly broken power law plus several Lorentzians gives similar $\chi^2$ values as our multi-Lorentzian fit. As pointed out by \citet{Belloni02}, the advantage of the multi-Lorentzian model fit is that it faciliates comparison across source types. Figure \ref{fig:pds} shows an example of the five-Lorentzian fits of the PDS. The errors are derived by varying the parameters until $\Delta\chi^2=1$, at 1 $\sigma$ level.

\section{Results}

\begin{figure}[htbp]
\centerline{\includegraphics[height=8.6cm,angle=-90]{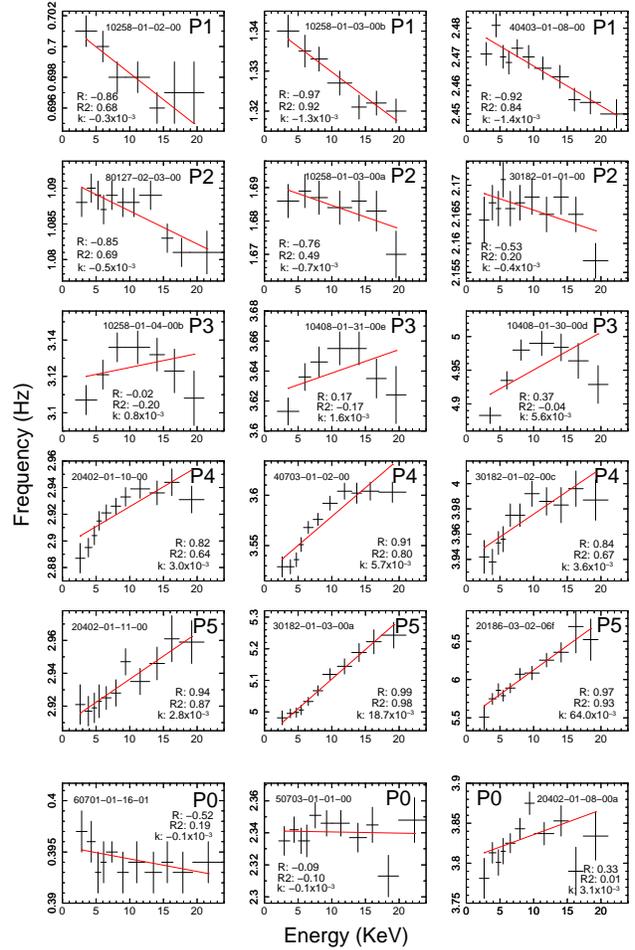}}
\caption{Three examples of frequency-energy relation for each pattern. The horizontal bars are band widths, and the vertical bars are error bars. The relations are fitted with Least Squares and the red oblique lines are the Best Fitting Lines. ``R" denotes correlation coefficient, ``R2" denotes adjusted R-square and ``k" denotes the slope of the fitting line.
P1: the QPO frequency decreases monotonously with energy, R $<$ -0.8, R2 $>$ 0.6;
P2: the frequency roughly maintains a constant at low energy and decreases with the energy at high energy;
P3: the relation evolves from a positive correlation to a negative one as the energy increases;
P4: the QPO frequency increases with energy at low energy and then it approximates a constant at high energy;
P5: the QPO frequency increases monotonously with energy, R $>$ 0.8, R2 $>$ 0.6.
P0: the irregular/indistinct relations.} 
\label{fig:e_fre} 
\end{figure}

For each observation interval listed in Table \ref{table1}, we have drawn a diagram that exhibits the relation between QPO frequency and photon energy. We use Least Squares \citep[see, e.g.,][]{Greene02} to estimate the linear correlation of the relations and obtain corresponding correlation coefficients (R), adjusted R-squares (R2) and slopes (k). Despite the complexity of the relations, they show some similar features. We find that it is possible to classify the relations into only six patterns, based on the appearance of the relation and the results of Least Squares fit. For the purpose of reducing the complexity of the amount of available data, we first focus on the relations with distinctive appearance and include them into corresponding patterns (see Table \ref{table1}). Figure \ref{fig:e_fre} shows three examples of energy-frequency relation for each pattern. In some cases, the QPO frequency decreases monotonously with energy, and R $<$ -0.8, R2 $>$ 0.6. We call this relation P1. Nevertheless, the relation is sometimes a positive ``linear'' correlation with R $>$ 0.8, R2 $>$ 0.6, which is referred to as P5. The other three patterns are the different combinations of P1 and P5. For P2, the frequency roughly maintains a constant at low energy and decreases with the energy at high energy. For P3, the relation evolves from a positive correlation (P5) to a negative one (P1) as the energy increases. While for P4, the frequency increases with energy at low energy and then it approximates a constant at high energy. The remaining relations are irregular/indistinct and hard to be included into certain patterns mentioned above. We call them P0. The P1 and P5 were firstly found by \citet{Qu10}, while P2, P3, P4 were newly discovered.

\begin{figure}[htbp]
\centerline{\includegraphics[height=8cm,angle=-90]{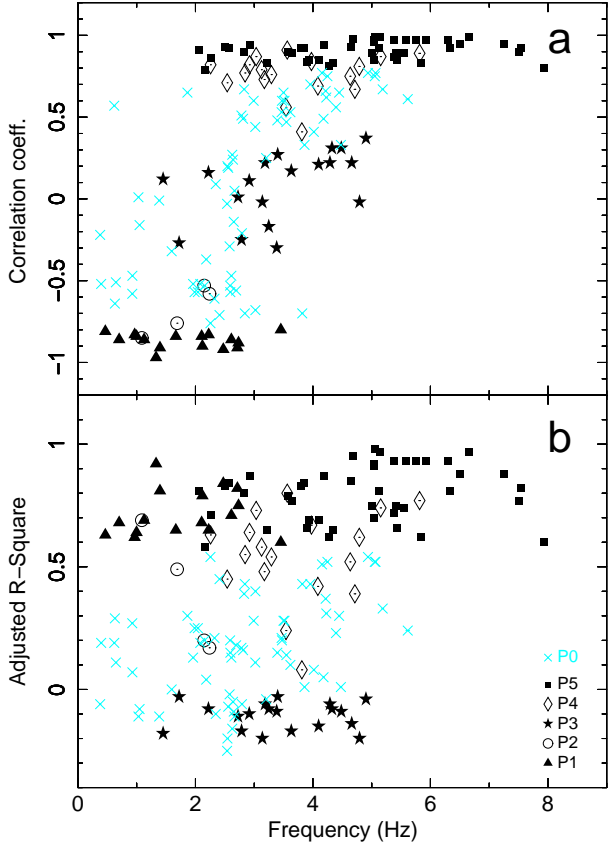}}
\caption{The QPO frequencies versus the correlation coefficients (a) and the adjusted R-squares (b) of six patterns. The triangles are observation intervals whose energy-frequency relations are belong to P1, the circles P2, the pentagons P3, the diamond P4, the squares P5, and the cyan crosses P0.
}
\label{fig:fre_cor} 
\end{figure}

The correlation coefficients and adjusted R-squares of each pattern are shown in Figure \ref{fig:fre_cor}. It is clear that P1 and P5 possess significant linear correlation, P3 non-linear correlation, while P2, P4 being mediate. The P0 points scatter among the points of P2, P3 and P4. In the order of P1 $\rightarrow$ P2 $\rightarrow$ P3 $\rightarrow$ P4 $\rightarrow$ P5, the correlation coefficients increase gradually from $\sim$ -1 to $\sim$ 1. This supports our classification as optimal.

\begin{figure}[htbp]
\centerline{\includegraphics[height=8cm,angle=-90]{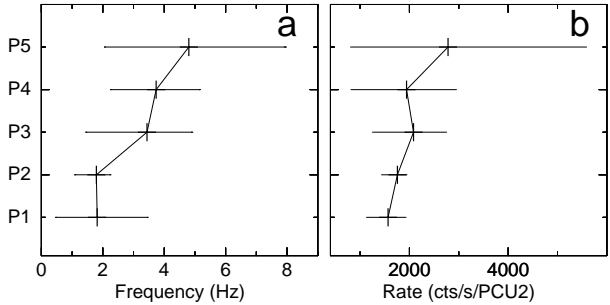}}
\caption{The QPO frequency statistics (a) and source count rate statistics (b) for five patterns of the relation between QPO frequency and photon energy.
The solid horizontal lines denotes the width of the band where QPO frequency or count rate measurements fall.
The crosses denotes the averaged QPO frequency or count rate of observation intervals of each pattern.}
\label{fig:shape_fre_rate} 
\end{figure}

Generally, with increasing of the QPO frequency and source intensity, the relation evolves from P1 to P5, via P2, P3 and P4, as demonstrated by Figure \ref{fig:shape_fre_rate}: the averaged QPO frequencies and intensities of P1 and P2 are obviously lower than those of P3, P4, and P5; the cases when QPO frequency $f\gtrsim 5$ Hz or count rate $\gtrsim 3000$ cts/s/PCU2 only occur in P5 and all the count rates and QPO frequencies of P1 and P2 are, respectively, less than $\sim$2000 cts/s/PCU2 and $\sim$4 Hz.   

\begin{figure}[htbp]
\centerline{\includegraphics[height=8cm,angle=-90]{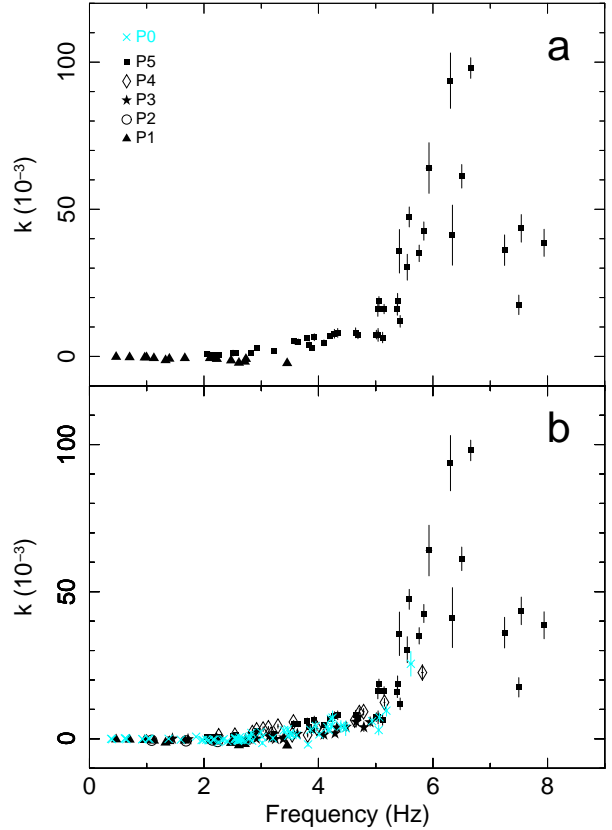}}
\caption{(a) the relation between QPO frequencies and the slopes of P1 and P5. 
(b) the relation between QPO frequencies and the slopes (k) of all (168) observations listed in Table \ref{table1}.
The horizontal and vertical bars are error bars. Some error bars are smaller than the symbols. The triangles are observation intervals whose energy-frequency relations are belong to P1, the circles P2, the pentagons P3, the diamond P4, the squares P5, and the cyan crosses P0.}
\label{fig:fre_k}
\end{figure}

The relation between QPO frequency and the slope is shown in Figure \ref{fig:fre_k}. Panel (a) of Figure \ref{fig:fre_k} demonstrates the slopes of P1 (k $<$ 0, $f < 3$ Hz) and P5 (k $>$ 0, $f > 2$ Hz). It is clearly that as QPO
frequency increases, the relation between QPO frequency and photon energy evolves from the negative correlation (P1) to the positive one (P5). Panel (b) of Figure \ref{fig:fre_k} demonstrates the slopes of all (168) observation intervals listed in Table \ref{table1}. The track in panel (b) is similar to that in panel (a), which confirms the evolution from P1 to P5, via P2, P3 and P4. This also suggests that our patterns are representative.

%\afterpage{\clearpage}

\section{Discussion and Conclusions}

By systematically analyzing all the {\it RXTE}/PCA observations of GRS 1915+105 before November 2010, we have found that 
there are five typical patterns of the relation between QPO frequency and photon energy: the negative correlation (P1) and 
the positive one (P5), as well as three combined relations of them (P2, P3 and P4). The P1 and P5 were firstly found by \citet{Qu10} and the intermediate patterns, P2, P3 and P4, were newly discovered in this work. Besides, we have confirmed in large sample the result reported by \citet{Qu10}: with increasing of the centroid frequency of QPO, the relation between QPO frequency and photon energy evolves from P1 to P5.

Following \citet{Qu10}, we apply several models to the detected patterns between the QPO frequency and photon energy. 
\citet{Titarchuk00} proposed the global disk model (GDM) to interpret the LFQPOs in X-ray binaries, in which they argued 
that the disk oscillations were the result of gravitational interaction between the the central compact objects and the disk and the QPO frequency was determined by the mass of the central object. Thus, the QPO frequency determined by the GDM 
is expected to be independent of the photon energy and therefore, this model can explain those QPOs whose coefficients are zero, as demonstrated in bottom panel of Figure \ref{fig:fre_k}. The radial and orbital oscillation model (ROOM) is a typical model for QPOs \citep{Nowak93, Nowak94}. According to this model, the oscillation frequency will vary with the the disk radius or temperature, resulting in that the QPO frequency will vary with the photon energy, because photons with different energies are from different radii. Simply assuming the disk oscillation frequency to be the Keplerian frequency at the inner disk radius, it is natural to explain the relation P5: with decreasing of the inner disk radius, the QPO frequency increases, and meanwhile the photon energy increases, because the temperature at inner disk edge also increases. Fortunately, \citet{Muno99} reported that in this source the QPO parameters were indeed correlated with the disk parameters. Other models such as the drift blob model \citep[DBM;][]{Bottcher98, Bottcher99} can also be used to explain the energy dependence of QPO frequency. In the DBM, the blobs drifting inward through an inhomogeneous hot inner disk or corona would cause the QPO frequency to increase with energy. Both the ROOM and DBM can explain the positive correlation between the QPO frequency and photon energy, but they are frustrated by the negative correlation (P1) and other complicated correlations (P2, P3, and P4). There is no model for interpreting all the relations, which perhaps indicates that the QPOs with different relations between the QPO frequency and photon energy result from different mechanisms.     

This work presents various patterns of relation between the QPO frequency and photon energy in the famous microquasar GRS 1915+105, indicating that the mechanisms for QPOs are complicated. On the other hands, we extend the investigation on the relations made by \citet{Qu10} and the newly discovered relations provide new clues on QPO models.  

\acknowledgments

The authors thank the anonymous referee for some helpful suggestions and comments. This research has made use of data obtained through the High Energy Astrophysics Science Archive Research Center (HEASARC) On-line Service, provided by NASA/ Goddard Space Flight Center (GSFC). This work is partially supported by the Natural Science Foundation of Xinjiang Uygur Autonomous Region of China (Grant No. 200821164), the Program of the Light in Chinese Western Region (LCWR) (Grant No. LHXZ 200802) provided by Chinese Academy of Sciences (CAS), the National Basic Research Program of China (973 Program 2009CB824800), the Natural Science Foundation of China (Grant No. 10733010, 11073021, 10325313, 10521001 and 10773017) and the Natural Science Foundation of China for Young Scientists (Grant No. 10903005).

\bibliography{references}
\bibliographystyle{Spr-mp-nameyear-cnd}

\clearpage

\begin{deluxetable}{lcrccccrccrccc}
\tabletypesize{\scriptsize\tiny}
\tablecolumns{14}
\tablewidth{0pc}
\tablecaption{The {\emph{RXTE}} Observations of GRS 1915+105 Reduced in This Paper and The Results}
\tablehead{\colhead{} & \colhead{} & \colhead{} & \colhead{} & \colhead{} & \multicolumn{3}{c}{QPO} & 
\colhead{} & \multicolumn{5}{c}{Energy--Frequency} \\
\cline{6-8} \cline{10-14} \\
\colhead{ObsID} & \colhead{Date} & \colhead{GTI$^a$} & \colhead{Rate$^b$} & \colhead{ChID$^c$} &
\colhead{Frequency} & \colhead{$\chi^2$} & \colhead{Q$^d$} & \colhead{ } &
\colhead{P$^e$} & \colhead{$k^f$} & \colhead{$\chi^2$} & \colhead{R$^g$}& \colhead{R2$^h$} \\
\colhead{} & \colhead{} & \colhead{(s)} & \colhead{} & \colhead{} &
\colhead{(Hz)} & \colhead{} & \colhead{} & \colhead{} &
\colhead{} & \colhead{($\times10^{-3}$)} & \colhead{} & \colhead{} & \colhead{}}
\startdata
10258-01-02-00 & 29/07/96& 9160&1739&Ch1E3& $0.699\pm0.001$& 2.16& $6.9\pm0.4$& &P1&  $-0.3\pm0.1$ & 0.65& -0.86& 0.68\\
10258-01-03-00a& 06/08/96& 3328&1757&Ch1E3& $1.687\pm0.004$& 3.22& $6.3\pm0.3$& &P2&  $-0.7\pm0.4$ & 0.52& -0.76& 0.49\\
10258-01-03-00b& 06/08/96& 3360&1771&Ch1E3& $1.329\pm0.003$& 2.08& $9.8\pm0.7$& &P1&  $-1.3\pm0.2$ & 0.52& -0.97& 0.92\\
10258-01-03-00c& 06/08/96& 3360&1736&Ch1E3& $1.450\pm0.004$& 1.82& $8.5\pm0.5$& &P3&  $-0.0\pm0.3$ & 1.55&  0.12&-0.18\\
10258-01-04-00a& 14/08/96& 6800&1915&Ch1E3& $2.692\pm0.003$& 1.87& $7.2\pm0.2$& &P0&  $-0.4\pm0.3$ & 1.85& -0.56& 0.18\\
10258-01-04-00b& 14/08/96& 3408&1971&Ch1E3& $3.129\pm0.007$& 2.14& $6.5\pm0.2$& &P3&  $ 0.8\pm0.7$ & 1.90& -0.02&-0.20\\
10258-01-05-00a& 20/08/96& 2688&3743&Ch2E3& $6.339\pm0.027$& 1.96& $3.4\pm0.2$& &P5& $41.2\pm10.2$ & 1.81&  0.92& 0.81\\
10258-01-05-00b& 20/08/96& 3376&3750&Ch2E3& $6.305\pm0.021$& 2.65& $3.8\pm0.2$& &P5&  $93.7\pm9.4$ & 0.47&  0.97& 0.93\\
10258-01-06-00a& 29/08/96& 1400&5549&Ch2E3& $7.250\pm0.040$& 1.28& $6.8\pm1.0$& &P5&  $36.1\pm5.2$ & 2.10&  0.95& 0.88\\
10258-01-06-00b& 29/08/96& 3408&5587&Ch2E3& $7.527\pm0.027$& 1.20& $7.5\pm1.0$& &P5&  $17.5\pm3.3$ & 1.71&  0.90& 0.77\\
10408-01-22-00 & 11/07/96& 3328&2122&Ch2E3& $3.473\pm0.005$& 0.88&$10.1\pm0.5$& &P0&  $ 1.0\pm0.6$ & 0.95&  0.60& 0.20\\
10408-01-22-01 & 11/07/96& 3312&2020&Ch2E3& $2.785\pm0.005$& 2.31& $7.1\pm0.2$& &P3&  $-0.1\pm0.6$ & 1.09& -0.25&-0.17\\
10408-01-22-02a& 11/07/96& 1600&1989&Ch2E3& $2.545\pm0.008$& 2.08& $7.7\pm0.5$& &P0&  $ 0.3\pm0.9$ & 0.46&  0.20&-0.20\\
10408-01-22-02b& 11/07/96&  820&1954&Ch2E3& $2.507\pm0.008$& 2.08& $8.5\pm0.8$& &P5&  $ 1.1\pm1.0$ & 0.05&  0.93& 0.83\\
10408-01-22-02c& 11/07/96&  892&1929&Ch2E3& $2.625\pm0.009$& 2.08& $8.2\pm0.6$& &P0&  $ 0.5\pm1.1$ & 0.48&  0.27&-0.16\\
10408-01-23-00a& 14/07/96& 3167&2109&Ch2E3& $3.499\pm0.006$& 1.90& $7.3\pm0.2$& &P0&  $ 1.7\pm0.7$ & 1.71&  0.65& 0.28\\
10408-01-23-00b& 14/07/96& 3312&2108&Ch2E3& $3.615\pm0.005$& 1.76& $8.7\pm0.3$& &P0&  $ 1.2\pm0.6$ & 1.50&  0.53& 0.10\\
10408-01-23-00c& 14/07/96& 3257&2255&Ch2E3& $4.179\pm0.008$& 1.50& $6.3\pm0.2$& &P0&  $ 3.0\pm1.0$ & 3.02&  0.49& 0.05\\
10408-01-24-00a& 16/07/96& 2447&1949&Ch2E3& $2.238\pm0.006$& 2.36& $8.1\pm0.5$& &P2&  $-0.9\pm0.5$ & 2.19& -0.58& 0.17\\
10408-01-24-00b& 16/07/96& 3312&1943&Ch2E3& $2.324\pm0.005$& 2.60& $7.2\pm0.3$& &P0&  $-0.5\pm0.5$ & 0.53& -0.61& 0.21\\
10408-01-24-00c& 16/07/96& 2953&1952&Ch2E3& $2.537\pm0.005$& 2.34& $9.4\pm0.5$& &P0&  $-0.0\pm0.5$ & 0.35& -0.03&-0.25\\
10408-01-24-00d& 16/07/96&  913&1965&Ch2E3& $2.594\pm0.008$& 1.59&$11.0\pm0.8$& &P0&  $-0.7\pm0.8$ & 0.57& -0.57& 0.15\\
10408-01-25-00 & 19/07/96& 9952&1820&Ch1E3& $1.126\pm0.002$& 3.34& $6.3\pm0.2$& &P1&  $-0.6\pm0.2$ & 1.45& -0.86& 0.69\\
10408-01-27-00a& 26/07/96& 2336&1783&Ch1E3& $0.645\pm0.002$& 1.43& $7.6\pm0.9$& &P0&  $-0.1\pm0.1$ & 0.42& -0.51& 0.11\\
10408-01-27-00b& 26/07/96& 3296&1791&Ch1E3& $0.617\pm0.002$& 1.20& $7.3\pm0.6$& &P0&  $ 0.1\pm0.2$ & 0.21&  0.57& 0.19\\
10408-01-27-00c& 26/07/96& 3296&1769&Ch1E3& $0.629\pm0.002$& 1.23& $6.1\pm0.5$& &P0&  $ 0.2\pm0.2$ & 0.26& -0.64& 0.29\\
10408-01-28-00a& 03/08/96& 3328&1742&Ch1E3& $0.996\pm0.002$& 1.60& $8.5\pm0.6$& &P1&  $-0.3\pm0.2$ & 0.53& -0.84& 0.64\\
10408-01-28-00b& 03/08/96& 3328&1744&Ch1E3& $0.965\pm0.004$& 2.14& $8.9\pm0.6$& &P1&  $-0.4\pm0.2$ & 0.22& -0.83& 0.62\\
10408-01-28-00c& 03/08/96& 3328&1731&Ch1E3& $0.927\pm0.003$& 1.24& $7.8\pm0.5$& &P0&  $-0.1\pm0.2$ & 0.33& -0.47& 0.07\\
10408-01-29-00a& 10/08/96& 2965&1760&Ch1E3& $1.665\pm0.004$& 1.66&$10.1\pm0.6$& &P1&  $-0.6\pm0.3$ & 0.33& -0.84& 0.65\\
10408-01-29-00b& 10/08/96& 3392&1784&Ch1E3& $1.863\pm0.004$& 2.38& $8.9\pm0.5$& &P0&  $ 0.7\pm0.3$ & 0.63&  0.65& 0.30\\
10408-01-29-00c& 10/08/96& 3392&1787&Ch1E3& $1.960\pm0.004$& 2.80& $9.4\pm0.5$& &P0&  $-0.4\pm0.3$ & 0.55& -0.52& 0.13\\
10408-01-30-00a& 18/08/96& 1696&2388&Ch1E3& $4.323\pm0.012$& 2.13& $5.8\pm0.3$& &P3&  $ 3.8\pm1.4$ & 4.68&  0.31&-0.08\\
10408-01-30-00b& 18/08/96& 1696&2588&Ch1E3& $4.792\pm0.011$& 1.36& $7.1\pm0.4$& &P3&  $ 3.9\pm1.4$ & 6.32& -0.02&-0.20\\
10408-01-30-00c& 18/08/96& 1696&2842&Ch1E3& $5.187\pm0.017$& 1.38& $6.2\pm0.4$& &P0&  $ 9.6\pm1.9$ & 2.44&  0.67& 0.33\\
10408-01-30-00d& 18/08/96& 1696&2752&Ch1E3& $4.901\pm0.011$& 1.06& $7.7\pm0.7$& &P3&  $ 5.6\pm1.3$ & 5.15&  0.37&-0.04\\
10408-01-30-00e& 18/08/96& 1688&2986&Ch1E3& $5.427\pm0.015$& 1.00& $8.6\pm0.7$& &P5&  $11.9\pm2.0$ & 1.92&  0.85& 0.66\\
10408-01-31-00a& 25/08/96& 2319&2327&Ch1E3& $4.096\pm0.006$& 1.15& $9.3\pm0.5$& &P3&  $ 1.4\pm0.6$ & 2.83&  0.21&-0.15\\
10408-01-31-00b& 25/08/96& 1000&2555&Ch1E3& $4.660\pm0.015$& 1.31& $8.2\pm0.7$& &P3&  $ 5.7\pm1.7$ & 3.93&  0.22&-0.14\\
10408-01-31-00c& 25/08/96& 1328&2496&Ch1E3& $4.482\pm0.014$& 1.30& $7.2\pm0.4$& &P3&  $ 4.0\pm1.4$ & 4.38&  0.31&-0.09\\
10408-01-31-00d& 25/08/96& 1000&2323&Ch1E3& $4.154\pm0.014$& 1.32& $8.2\pm0.7$& &P0&  $ 4.0\pm1.3$ & 1.25&  0.77& 0.51\\
10408-01-31-00e& 25/08/96& 1664&2133&Ch1E3& $3.630\pm0.009$& 1.58& $7.4\pm0.4$& &P3&  $ 1.6\pm0.8$ & 2.07&  0.17&-0.17\\
10408-01-31-00f& 25/08/96& 1664&2057&Ch1E3& $3.382\pm0.008$& 1.52& $9.7\pm0.5$& &P3&  $-0.1\pm0.7$ & 3.09& -0.30&-0.09\\
10408-01-32-00a& 31/08/96& 2912&4239&Ch1E3& $6.503\pm0.026$& 2.53& $4.6\pm0.6$& &P5&  $61.2\pm4.0$ & 2.33&  0.95& 0.88\\
10408-01-32-00b& 31/08/96& 3312&3648&Ch1E3& $5.840\pm0.017$& 2.19& $7.0\pm0.7$& &P5&  $42.6\pm3.1$ & 2.65&  0.83& 0.62\\
10408-01-32-00c& 31/08/96& 1170&3314&Ch1E3& $5.613\pm0.032$& 1.17& $5.7\pm0.5$& &P0&  $25.5\pm4.2$ & 2.31&  0.61& 0.24\\
10408-01-33-00a& 07/09/96&  912&3527&Ch1E3& $5.550\pm0.030$& 1.67& $5.2\pm0.6$& &P5&  $30.3\pm4.4$ & 1.30&  0.89& 0.74\\
10408-01-33-00b& 07/09/96& 2495&3743&Ch1E3& $5.586\pm0.017$& 1.68& $7.4\pm0.8$& &P5&  $47.4\pm3.4$ & 2.87&  0.97& 0.93\\
10408-01-33-00c& 07/09/96& 1295&3655&Ch1E3& $5.414\pm0.039$& 1.91& $3.8\pm0.4$& &P5&  $35.7\pm7.4$ & 0.23&  0.89& 0.75\\
10408-01-42-00a& 23/10/96& 3312&3289&Ch1E3& $5.044\pm0.009$& 2.49& $8.2\pm0.4$& &P5&  $ 7.4\pm1.3$ & 1.08&  0.97& 0.92\\
10408-01-42-00b& 23/10/96& 3312&2921&Ch1E3& $4.690\pm0.009$& 2.45& $7.1\pm0.3$& &P5&  $ 7.1\pm1.1$ & 0.73&  0.98& 0.95\\
10408-01-43-00a& 23/10/96& 2416&3274&Ch1E3& $5.011\pm0.010$& 2.14& $8.3\pm0.5$& &P5&  $ 7.4\pm1.5$ & 2.09&  0.89& 0.75\\
10408-01-43-00b& 23/10/96& 2284&3314&Ch1E3& $5.064\pm0.012$& 1.86& $8.1\pm0.4$& &P0&  $ 7.3\pm1.5$ & 2.99&  0.77& 0.52\\
10408-01-43-00c& 23/10/96& 1980&3302&Ch1E3& $5.121\pm0.013$& 1.90& $7.3\pm0.4$& &P5&  $ 6.4\pm1.8$ & 1.44&  0.92& 0.81\\
10408-01-43-00d& 23/10/96& 1740&2709&Ch1E3& $4.446\pm0.013$& 1.43& $6.7\pm0.4$& &P0&  $ 4.2\pm1.5$ & 0.73&  0.65& 0.30\\
20186-03-02-052a&17/09/97& 3031&3096&Ch3E3& $5.381\pm0.019$& 1.63& $7.5\pm0.7$& &P5&  $18.8\pm2.6$ & 1.82&  0.87& 0.72\\
20186-03-02-052b&17/09/97& 3031&3203&Ch3E3& $5.759\pm0.018$& 2.11& $4.7\pm0.2$& &P5&  $35.0\pm2.8$ & 2.27&  0.97& 0.93\\
20186-03-02-052c&17/09/97& 3312&2348&Ch3E3& $4.083\pm0.006$& 1.73& $7.9\pm0.3$& &P4&  $ 2.6\pm0.6$ & 2.73&  0.69& 0.42\\
20186-03-02-052d&17/09/97& 3312&2563&Ch3E3& $4.633\pm0.008$& 1.37& $6.8\pm0.2$& &P4&  $ 6.4\pm0.9$ & 3.09&  0.75& 0.52\\
20186-03-02-052e&17/09/97& 3312&2802&Ch3E3& $5.153\pm0.011$& 1.52& $7.3\pm0.4$& &P4&  $12.4\pm1.3$ & 3.44&  0.87& 0.74\\
20186-03-02-060a&18/09/97& 2768&2852&Ch3E3& $5.036\pm0.021$& 2.07& $4.8\pm0.3$& &P5&  $16.1\pm2.5$ & 0.79&  0.96& 0.91\\
20186-03-02-060b&18/09/97& 9936&4385&Ch3E3& $6.658\pm0.031$& 2.26& $2.4\pm0.1$& &P5&  $98.0\pm3.5$ & 1.92&  0.99& 0.97\\
20186-03-02-060c&18/09/97& 3312&2679&Ch3E3& $4.788\pm0.011$& 1.34& $5.7\pm0.2$& &P4&  $ 9.2\pm1.1$ & 3.99&  0.81& 0.62\\
20186-03-02-06a& 18/09/97& 1656&2767&Ch3E3& $5.042\pm0.017$& 1.17& $5.1\pm0.3$& &P5&  $ 7.3\pm2.2$ & 0.65&  0.85& 0.70\\
20186-03-02-06b& 18/09/97& 1656&2430&Ch3E3& $4.226\pm0.013$& 1.33& $7.1\pm0.3$& &P0&  $ 4.3\pm1.1$ & 2.72&  0.62& 0.31\\
20186-03-02-06c& 18/09/97& 1600&2255&Ch3E3& $3.810\pm0.008$& 1.13& $8.2\pm0.4$& &P4&  $ 1.3\pm0.7$ & 1.69&  0.41& 0.08\\
20186-03-02-06d& 18/09/97& 1695&2399&Ch3E3& $4.288\pm0.009$& 1.49& $7.8\pm0.4$& &P3&  $ 1.8\pm0.9$ & 1.89&  0.22&-0.06\\ 
20186-03-02-06e& 18/09/97& 1550&2761&Ch3E3& $5.051\pm0.013$& 0.70& $7.3\pm0.5$& &P0&  $ 3.0\pm1.7$ & 0.56&  0.75& 0.52\\
20186-03-02-06f& 18/09/97& 1569&3648&Ch3E3& $5.930\pm0.050$& 1.35& $2.6\pm0.2$& &P5&  $64.0\pm8.6$ & 0.40&  0.97& 0.93\\
20402-01-05-00 & 05/12/96& 2048&1421&Ch5E3& $2.825\pm0.004$& 1.89& $6.1\pm0.2$& &P5&  $ 1.2\pm0.3$ & 0.34&  0.90& 0.80\\
20402-01-06-00a& 11/12/96& 3312&1360&Ch5E3& $3.034\pm0.009$& 1.00& $5.0\pm0.3$& &P4&  $ 3.3\pm0.8$ & 0.77&  0.87& 0.73\\
20402-01-06-00b& 11/12/96& 3312&1279&Ch5E3& $2.844\pm0.007$& 1.31& $5.6\pm0.3$& &P4&  $ 2.1\pm0.6$ & 1.46&  0.77& 0.55\\
20402-01-06-00c& 11/12/96& 2780&1211&Ch5E3& $2.568\pm0.007$& 1.42& $5.7\pm0.3$& &P5&  $ 1.2\pm0.6$ & 0.08&  0.92& 0.84\\
20402-01-07-00 & 19/12/96& 9296&1310&Ch5E3& $3.125\pm0.005$& 1.48& $4.2\pm0.1$& &P4&  $ 3.7\pm0.5$ & 2.54&  0.79& 0.58\\
20402-01-08-00a& 24/12/96& 2658&1318&Ch5E3& $3.845\pm0.010$& 1.63& $5.1\pm0.2$& &P0&  $ 3.1\pm1.3$ & 2.09&  0.33& 0.01\\
20402-01-08-00b& 24/12/96& 2834&1325&Ch5E3& $3.927\pm0.010$& 2.27& $5.0\pm0.2$& &P5&  $ 6.5\pm1.2$ & 1.36&  0.85& 0.69\\
20402-01-08-01 & 25/12/96& 3312&1232&Ch5E3& $3.465\pm0.009$& 1.25& $4.4\pm0.2$& &P0&  $ 3.2\pm1.1$ & 1.54&  0.53& 0.21\\
20402-01-09-00 & 31/12/96& 7548&1099&Ch5E3& $2.827\pm0.005$& 1.87& $5.1\pm0.2$& &P0&  $ 2.0\pm0.5$ & 1.46&  0.67& 0.39\\
20402-01-10-00 & 08/01/97& 9804& 993&Ch5E3& $2.920\pm0.005$& 2.22& $4.8\pm0.2$& &P4&  $ 3.0\pm0.5$ & 1.84&  0.82& 0.64\\
20402-01-11-00 & 14/01/97& 6519& 912&Ch5E3& $2.930\pm0.006$& 1.33& $5.2\pm0.2$& &P5&  $ 2.8\pm0.7$ & 0.42&  0.94& 0.87\\
20402-01-12-00a& 23/01/97& 5695& 883&Ch5E3& $2.811\pm0.006$& 1.60& $5.3\pm0.3$& &P0&  $ 0.9\pm0.7$ & 0.86&  0.49& 0.16\\
20402-01-12-00b& 23/01/97& 3755& 894&Ch5E3& $2.790\pm0.007$& 1.50& $6.0\pm0.4$& &P0&  $ 0.0\pm0.8$ & 0.91& -0.21&-0.06\\
20402-01-13-00 & 29/01/97&10000& 936&Ch5E3& $3.649\pm0.007$& 1.57& $4.0\pm0.1$& &P5&  $ 5.0\pm0.9$ & 1.64&  0.89& 0.77\\
20402-01-14-00 & 01/02/97& 9394& 910&Ch5E3& $3.577\pm0.007$& 2.58& $4.0\pm0.1$& &P5&  $ 5.1\pm0.9$ & 1.46&  0.90& 0.79\\
20402-01-15-00 & 09/02/97&10222& 816&Ch5E3& $2.258\pm0.004$& 2.14& $5.9\pm0.2$& &P4&  $ 1.0\pm0.4$ & 0.43&  0.82& 0.63\\
20402-01-16-00 & 22/02/97& 5951& 803&Ch5E3& $2.991\pm0.007$& 1.49& $5.4\pm0.3$& &P0&  $ 2.3\pm0.7$ & 2.59&  0.60& 0.28\\
20402-01-20-00 & 17/03/97& 7300& 807&Ch5E3& $3.217\pm0.006$& 1.19& $5.4\pm0.2$& &P5&  $ 1.7\pm0.7$ & 0.72&  0.83& 0.65\\
20402-01-26-00a& 25/04/97& 2220&1137&Ch5E3& $3.954\pm0.012$& 1.47& $4.9\pm0.3$& &P0&  $ 4.4\pm1.6$ & 1.13&  0.70& 0.43\\
20402-01-26-00b& 25/04/97& 2884&1188&Ch5E3& $4.279\pm0.011$& 2.19& $4.7\pm0.2$& &P5&  $ 7.7\pm1.5$ & 1.52&  0.81& 0.62\\
20402-01-26-00c& 25/04/97& 3300&1210&Ch5E3& $4.475\pm0.016$& 1.82& $3.6\pm0.2$& &P0&  $ 3.6\pm2.5$ & 0.77&  0.33& 0.01\\
20402-01-26-00d& 25/04/97& 3328&1178&Ch5E3& $4.246\pm0.017$& 1.68& $3.5\pm0.2$& &P0&  $ 7.2\pm2.3$ & 0.66&  0.75& 0.52\\
20402-01-26-00e& 25/04/97& 1964&1163&Ch5E3& $4.391\pm0.014$& 2.08& $4.8\pm0.2$& &P0&  $ 4.2\pm2.0$ & 1.77&  0.56& 0.23\\
20402-01-48-00a& 29/09/97& 3296&4714&Ch5E3& $7.541\pm0.038$& 1.54& $5.2\pm0.4$& &P5&  $43.5\pm4.7$ & 1.99&  0.92& 0.82\\
20402-01-48-00b& 29/09/97& 3328&2726&Ch5E3& $4.713\pm0.014$& 1.73& $5.7\pm0.4$& &P4&  $ 8.9\pm1.6$ & 3.93&  0.67& 0.39\\
20402-01-50-01 & 16/10/97& 4994&1497&Ch5E3& $1.048\pm0.003$& 2.39& $8.2\pm0.5$& &P0&  $-0.0\pm0.2$ & 0.40& -0.16&-0.08\\
20402-01-51-00 & 22/10/97& 9399&1490&Ch5E3& $1.396\pm0.002$& 3.47& $7.8\pm0.2$& &P1&  $-0.8\pm0.1$ & 0.71& -0.91& 0.81\\
30182-01-01-00 & 08/07/98&11606&1435&Ch3E3& $2.148\pm0.003$& 2.36& $5.7\pm0.2$& &P2&  $-0.4\pm0.2$ & 0.97& -0.53& 0.20\\
30182-01-02-00a& 09/07/98& 5073&1889&Ch3E3& $3.247\pm0.005$& 1.94& $7.7\pm0.3$& &P3&  $-0.1\pm0.4$ & 0.62& -0.17&-0.08\\
30182-01-02-00b& 09/07/98& 3359&2069&Ch3E3& $3.541\pm0.006$& 2.23& $7.3\pm0.3$& &P4&  $ 1.4\pm0.6$ & 1.95&  0.56& 0.24\\
30182-01-02-00c& 09/07/98& 2968&2466&Ch3E3& $3.973\pm0.008$& 2.89& $6.0\pm0.3$& &P4&  $ 3.6\pm0.8$ & 1.17&  0.84& 0.67\\
30182-01-03-00a& 10/07/98& 3344&3479&Ch3E3& $5.056\pm0.013$& 2.19& $6.4\pm0.3$& &P5&  $18.7\pm1.6$ & 0.43&  0.99& 0.98\\
30182-01-03-00b& 10/07/98& 2472&3677&Ch3E3& $5.147\pm0.010$& 2.77& $9.3\pm0.6$& &P5&  $16.1\pm1.6$ & 0.69&  0.99& 0.97\\
30182-01-04-00a& 11/07/98& 1678&2360&Ch3E3& $4.096\pm0.010$& 1.46& $7.0\pm0.5$& &P5&  $ 4.6\pm1.0$ & 1.03&  0.85& 0.69\\
30182-01-04-00b& 11/07/98& 4166&1933&Ch3E3& $3.400\pm0.006$& 2.18& $7.4\pm0.3$& &P3&  $ 0.4\pm0.5$ & 0.98&  0.27&-0.03\\
30182-01-04-00c& 11/07/98& 3328&1709&Ch3E3& $2.916\pm0.006$& 2.25& $9.1\pm0.4$& &P3&  $ 0.0\pm0.4$ & 1.21&  0.11&-0.10\\
30182-01-04-00d& 11/07/98& 3324&1604&Ch3E3& $2.664\pm0.005$& 1.87& $9.0\pm0.4$& &P0&  $-0.2\pm0.4$ & 1.55&  0.05&-0.11\\
30182-01-04-01a& 12/07/98& 2236&1581&Ch3E3& $2.652\pm0.006$& 1.44& $8.8\pm0.5$& &P0&  $-0.3\pm0.4$ & 1.09& -0.14&-0.09\\
30182-01-04-01b& 12/07/98& 2728&1513&Ch3E3& $2.410\pm0.005$& 1.61& $7.6\pm0.3$& &P0&  $-0.8\pm0.4$ & 0.52& -0.71& 0.45\\
30182-01-04-01c& 12/07/98& 3340&1605&Ch3E3& $2.725\pm0.006$& 1.70& $7.4\pm0.3$& &P3&  $-0.1\pm0.5$ & 0.83&  0.01&-0.11\\
30182-01-04-01d& 12/07/98& 3340&2010&Ch3E3& $3.392\pm0.014$& 2.64& $4.3\pm0.2$& &P0&  $ 2.9\pm1.2$ & 1.51&  0.48& 0.15\\
30182-01-04-01e& 12/07/98& 2400&2665&Ch3E3& $4.215\pm0.010$& 2.74& $6.6\pm0.3$& &P0&  $ 6.8\pm1.1$ & 4.76&  0.66& 0.37\\
30402-01-09-01 & 10/04/98& 2546&1979&Ch6E3& $2.157\pm0.004$& 2.84& $8.9\pm0.4$& &P0&  $-0.3\pm0.3$ & 0.42& -0.53& 0.20\\
30402-01-10-00a& 11/04/98& 3312&1970&Ch6E3& $1.595\pm0.003$& 1.76& $8.3\pm0.4$& &P0&  $-0.3\pm0.3$ & 1.02& -0.32& 0.00\\
30402-01-10-00b& 11/04/98& 6303&1956&Ch6E3& $1.721\pm0.003$& 3.86& $8.6\pm0.3$& &P3&  $-0.2\pm0.2$ & 2.25& -0.27&-0.03\\
30402-01-11-00a& 20/04/98& 3311&2777&Ch6E3& $5.378\pm0.013$& 1.92& $4.2\pm0.2$& &P5&  $16.0\pm2.0$ & 0.77&  0.97& 0.93\\
30402-01-11-00b& 20/04/98& 2271&2952&Ch6E3& $5.815\pm0.017$& 1.75& $4.2\pm0.2$& &P4&  $22.5\pm2.7$ & 2.23&  0.89& 0.77\\
30703-01-16-00 & 28/04/98& 5038&1816&Ch6E3& $1.382\pm0.003$& 2.76& $7.1\pm0.3$& &P0&  $-0.0\pm0.2$ & 0.36& -0.01&-0.11\\
30703-01-17-00 & 06/05/98& 4584&1739&Ch6E3& $0.925\pm0.002$& 1.00& $8.2\pm0.5$& &P0&  $-0.1\pm0.1$ & 0.26& -0.58& 0.27\\
30703-01-22-00 & 27/06/98& 3375&1539&Ch6E3& $2.256\pm0.005$& 1.79& $7.5\pm0.3$& &P0&  $-1.1\pm0.4$ & 0.68& -0.76& 0.54\\
30703-01-25-00a& 23/07/98& 2626&1718&Ch6E3& $3.172\pm0.007$& 1.08& $7.1\pm0.4$& &P4&  $ 1.5\pm0.6$ & 0.76&  0.73& 0.48\\
30703-01-25-00b& 23/07/98& 2322&2146&Ch6E3& $3.804\pm0.009$& 1.36& $5.7\pm0.3$& &P5&  $ 6.2\pm0.9$ & 1.15&  0.92& 0.83\\
30703-01-33-00 & 15/09/98& 4917&1400&Ch6E3& $3.293\pm0.007$& 1.07& $4.2\pm0.2$& &P4&  $ 4.3\pm0.7$ & 4.21&  0.76& 0.54\\
30703-01-41-00 & 26/12/98& 4707&1233&Ch6E3& $2.158\pm0.004$& 1.59& $8.7\pm0.4$& &P5&  $ 0.5\pm0.3$ & 0.23&  0.79& 0.58\\
40403-01-08-00 & 02/06/99& 9884&1584&Ch6E4& $2.469\pm0.003$& 3.28& $7.7\pm0.2$& &P1&  $-1.4\pm0.2$ & 0.92& -0.92& 0.84\\
40403-01-09-00 & 08/07/99&13355&1343&Ch6E4& $2.041\pm0.003$& 2.95& $6.6\pm0.2$& &P0&  $-0.3\pm0.2$ & 0.97& -0.57& 0.25\\
40403-01-11-00 & 28/02/00&13355&2426&Ch6E4& $4.336\pm0.011$& 2.88& $6.3\pm0.3$& &P5&  $ 8.0\pm1.5$ & 0.87&  0.83& 0.65\\
40703-01-01-00 & 01/01/99& 9731&1281&Ch6E4& $2.265\pm0.003$& 1.84& $7.0\pm0.2$& &P5&  $ 0.6\pm0.2$ & 0.24&  0.86& 0.71\\
40703-01-02-00 & 08/01/99& 9005&1861&Ch6E4& $3.562\pm0.005$& 2.07& $4.7\pm0.1$& &P4&  $ 5.7\pm0.5$ & 3.19&  0.91& 0.80\\
40703-01-05-00 & 12/02/99&10129&1592&Ch6E4& $4.194\pm0.005$& 3.17& $4.7\pm0.1$& &P5&  $ 6.9\pm0.7$ & 1.72&  0.94& 0.87\\
40703-01-09-00 & 28/03/99& 4702&1418&Ch6E4& $2.782\pm0.005$& 1.18& $6.5\pm0.2$& &P0&  $ 0.5\pm0.4$ & 0.46&  0.51& 0.17\\
40703-01-38-02 & 15/11/99& 2501&5138&Ch6E4& $7.940\pm0.034$& 1.45& $5.2\pm0.5$& &P5&  $38.6\pm4.6$ & 2.18&  0.80& 0.60\\
50125-01-01-03 & 13/07/00& 2735&1747&Ch3E5& $3.019\pm0.006$& 1.98& $8.3\pm0.4$& &P0&  $ 0.7\pm0.4$ & 1.06&  0.44& 0.11\\
50125-01-03-00a& 15/07/00& 4348&2077&Ch3E5& $3.547\pm0.007$& 1.77& $5.9\pm0.2$& &P0&  $ 1.3\pm0.6$ & 1.50&  0.47& 0.14\\
50125-01-03-00b& 15/07/00&10652&1818&Ch3E5& $3.182\pm0.004$& 5.37& $6.5\pm0.2$& &P3&  $ 0.4\pm0.3$ & 2.82&  0.22&-0.06\\
50703-01-01-00 & 08/03/00& 4755&1314&Ch6E4& $2.345\pm0.007$& 1.34& $4.8\pm0.2$& &P0&  $-0.1\pm0.6$ & 0.99&  0.09&-0.10\\
50703-01-49-00 & 27/02/01& 5467&1434&Ch6E5& $2.610\pm0.004$& 2.77& $8.1\pm0.3$& &P0&  $-0.4\pm0.3$ & 0.53& -0.47& 0.13\\
50703-01-55-01 & 17/04/01& 6896&1583&Ch6E5& $2.840\pm0.004$& 2.51& $8.4\pm0.3$& &P0&  $-0.6\pm0.3$ & 0.77& -0.70& 0.43\\
50703-01-67-00 & 22/07/01& 1806&1243&Ch6E5& $2.182\pm0.005$& 1.57&$10.0\pm0.7$& &P0&  $-0.2\pm0.4$ & 0.17& -0.37& 0.04\\
60100-01-01-00 & 05/08/01& 3280&1249&Ch6E5& $2.226\pm0.004$& 1.50&$12.2\pm0.6$& &P1&  $-0.9\pm0.3$ & 0.70& -0.83& 0.65\\
60100-01-02-000a&06/08/01& 2748&1487&Ch6E5& $2.714\pm0.005$& 1.29& $8.3\pm0.4$& &P1&  $-1.8\pm0.4$ & 0.47& -0.91& 0.82\\
60100-01-02-000b&06/08/01& 2496&1654&Ch6E5& $3.021\pm0.006$& 1.09& $7.2\pm0.4$& &P0&  $-1.4\pm0.6$ & 0.87& -0.68& 0.40\\
60100-01-02-000c&06/08/01& 2648&1762&Ch6E5& $3.203\pm0.007$& 1.28& $6.6\pm0.3$& &P0&  $ 0.2\pm0.7$ & 0.49&  0.25&-0.04\\
60100-01-02-000d&06/08/01& 2816&1967&Ch6E5& $3.515\pm0.007$& 1.49& $6.7\pm0.3$& &P0&  $ 1.9\pm0.7$ & 2.04&  0.60& 0.28\\
60100-01-02-000e&06/08/01& 2964&2178&Ch6E5& $3.839\pm0.008$& 2.40& $6.4\pm0.3$& &P5&  $ 4.0\pm0.9$ & 0.62&  0.92& 0.84\\
60405-01-03-00 & 05/08/01& 6560&1474&Ch6E5& $2.732\pm0.004$& 1.64& $8.0\pm0.3$& &P1&  $-0.9\pm0.3$ & 0.36& -0.88& 0.75\\
60701-01-16-00 & 28/02/02& 3068&1820&Ch6E5& $0.377\pm0.002$& 0.75& $7.2\pm0.8$& &P0&  $-0.0\pm0.1$ & 0.18& -0.22&-0.06\\
60701-01-16-01 & 28/02/02& 3109&1809&Ch6E5& $0.395\pm0.002$& 0.90& $7.6\pm1.2$& &P0&  $-0.1\pm0.1$ & 0.37& -0.52& 0.19\\
60701-01-23-00 & 22/01/02& 3263&1986&Ch6E5& $2.082\pm0.003$& 3.92&$10.3\pm0.4$& &P0&  $-0.5\pm0.3$ & 1.33& -0.52& 0.19\\
60701-01-28-00 & 06/03/02& 9680&1744&Ch6E5& $0.466\pm0.001$& 0.99& $7.6\pm0.6$& &P1&  $-0.2\pm0.1$ & 0.76& -0.81& 0.63\\
60701-01-33-00 & 24/04/02& 3247&1426&Ch6E5& $1.029\pm0.002$& 1.24&$10.1\pm0.8$& &P0&  $ 0.0\pm0.1$ & 0.60&  0.01&-0.11\\
70702-01-23-00 & 03/10/02& 3231&1931&Ch6E5& $3.453\pm0.005$& 1.48& $9.8\pm0.4$& &P1&  $-2.3\pm0.5$ & 1.16& -0.80& 0.60\\
70702-01-24-00 & 09/10/02& 3264&1328&Ch6E5& $2.581\pm0.005$& 2.93&$13.2\pm0.7$& &P0&  $-0.3\pm0.3$ & 0.89& -0.29&-0.02\\
70703-01-01-08 & 01/04/02&10704&1902&Ch3E5& $2.589\pm0.004$& 5.81& $7.5\pm0.2$& &P0&  $-0.5\pm0.3$ & 2.75& -0.53& 0.20\\
70703-01-01-14 & 29/03/02& 8240&1869&Ch6E5& $2.639\pm0.004$& 5.81& $6.2\pm0.1$& &P0&  $-0.4\pm0.3$ & 2.36&  0.24&-0.05\\
80127-02-03-00 & 10/04/03&11728&1884&Ch4E5& $1.088\pm0.002$& 2.84& $6.9\pm0.3$& &P2&  $-0.5\pm0.1$ & 0.86& -0.85& 0.69\\
80701-01-08-00 & 25/10/06& 3216&2375&Ch6E5& $4.648\pm0.012$& 1.45& $6.3\pm0.3$& &P5&  $ 8.0\pm1.7$ & 0.43&  0.93& 0.85\\
80701-01-26-00 & 28/11/06& 6304&1334&Ch6E5& $2.541\pm0.004$& 2.72&$10.4\pm0.4$& &P4&  $ 1.2\pm0.3$ & 1.13&  0.71& 0.45\\
80701-01-32-00 & 04/12/06& 1239&1212&Ch6E5& $2.104\pm0.003$& 1.41&$10.3\pm0.4$& &P1&  $-0.6\pm0.2$ & 0.41& -0.84& 0.68\\
80701-01-51-00 & 09/12/06& 6960&1252&Ch6E5& $2.221\pm0.003$& 2.14& $9.4\pm0.4$& &P3&  $ 0.3\pm0.2$ & 1.74&  0.16&-0.08\\
80701-01-55-02 & 11/01/07& 5440&1131&Ch6E5& $2.611\pm0.004$& 2.16& $9.2\pm0.4$& &P1&  $-2.2\pm0.3$ & 2.28& -0.86& 0.71\\
80701-01-56-00 & 18/01/07& 9600&1073&Ch6E5& $2.558\pm0.001$& 2.39&$16.8\pm0.6$& &P0&  $ 0.0\pm0.1$ & 0.52&  0.19&-0.07\\
80701-01-57-00 & 24/01/07& 9584&1100&Ch6E5& $2.060\pm0.003$& 4.36&$13.8\pm0.5$& &P5&  $ 0.7\pm0.2$ & 0.31&  0.91& 0.81\\
90105-01-03-01 & 15/05/04& 7152&3023&Ch4E5& $4.936\pm0.009$& 1.93& $5.7\pm0.2$& &P0&  $ 6.0\pm1.1$ & 1.74&  0.77& 0.54\\
90105-07-01-00 & 12/04/05& 6464&2098&Ch4E5& $4.015\pm0.005$& 1.46& $7.9\pm0.2$& &P0&  $ 1.5\pm0.6$ & 1.45&  0.41& 0.08\\
90105-07-02-00 & 13/04/05& 6368&2123&Ch4E5& $3.895\pm0.006$& 2.94& $6.8\pm0.2$& &P5&  $ 3.0\pm0.6$ & 2.12&  0.84& 0.66\\
90701-01-19-00 & 28/07/04& 6416&1200&Ch6E5& $2.117\pm0.002$& 3.06&$13.2\pm0.5$& &P1&  $-0.4\pm0.1$ & 0.31& -0.90& 0.79\\
91701-01-55-00 & 02/05/07& 9584&1091&Ch6E5& $1.986\pm0.002$& 2.66&$23.7\pm1.0$& &P0&  $-0.4\pm0.2$ & 0.90& -0.57& 0.25\\
92702-01-09-00 & 04/05/06& 5136&1071&Ch6E5& $3.817\pm0.006$& 1.17& $7.6\pm0.3$& &P0&  $-1.9\pm0.7$ & 1.03& -0.70& 0.43\\
\enddata\label{table1}

\tablenotetext{a}{: Good time interval;}
\tablenotetext{a}{: The unit is cts/s/PCU2;}
\tablenotetext{c}{: ChID is sign which represent the definition of PCA energy bands for 
light curve extraction listed on Table \ref{table2};}
\tablenotetext{d}{: Coherence factor of QPO;}
\tablenotetext{e}{: Pattern of the relation between QPO frequency and photon energy;}
\tablenotetext{f}{: The Least Squares slope;}
\tablenotetext{g}{: Correlation coefficient;}
\tablenotetext{h}{: Adjusted R-square.}

\end{deluxetable}

\begin{deluxetable}{cccccccccccc}
\tabletypesize{\scriptsize\tiny}
\tablecolumns{12}
\tablewidth{0pc}
\tablecaption{The Definitions of {{\it RXTE}/PCA} Energy Bands for Light Curve Extraction.}
\tablehead{\colhead{ChID$^a$} & \colhead{Channel} & \colhead{Energy} & \colhead{Centroid} & 
\colhead{ChID$^a$} & \colhead{Channel} & \colhead{Energy} & \colhead{Centroid} & 
\colhead{ChID$^a$} & \colhead{Channel} & \colhead{Energy} & \colhead{Centroid} \\
\colhead{} & \colhead{} & \colhead{(keV)} & \colhead{(keV)} & \colhead{} & \colhead{} & 
\colhead{(keV)} & \colhead{(keV)} & \colhead{} & \colhead{} & \colhead{(keV)} & \colhead{(keV)}}
\startdata
Ch1E3   & 0-13 &  1.94-5.12 &  3.53&  Ch2E3 &  0-13 &  1.94-5.12 &  3.53&   Ch3E3  &  0-8  &  1.94-3.35 &  2.65 \\
        &14-18 &  5.12-6.89 &  6.01& 	    & 14-18 &  5.12-6.89 &  6.01&  	   &  9-11 &  3.35-4.41 &  3.88 \\
        &19-25 &  6.89-9.39 &  8.14& 	    & 19-25 &  6.89-9.39 &  8.14&  	   & 12-13 &  4.41-5.12 &  4.77 \\
        &26-35 &  9.39-12.99& 11.19& 	    & 26-35 &  9.39-12.99& 11.19&  	   & 14-15 &  5.12-5.82 &  5.47 \\
        &36-41 & 12.99-15.17& 14.08& 	    & 36-41 & 12.99-15.17& 14.08&  	   & 16-19 &  5.82-7.25 &  6.54 \\
        &42-49 & 15.17-18.09& 16.63& 	    & 42-49 & 15.17-18.09& 16.63&  	   & 20-23 &  7.25-8.68 &  7.97 \\
        &50-58 & 18.09-21.04& 19.57& 	    &       &            &      &  	   & 24-29 &  8.68-10.83&  9.76 \\
        &      &            &      & 	    &	    &		 &	&  	   & 30-35 & 10.83-12.99& 11.91 \\
	&      & 	    &	   & 	    &	    &		 &	&  	   & 36-41 & 12.99-15.17& 14.08 \\
	&      & 	    &	   & 	    &	    &		 &	&  	   & 42-47 & 15.17-17.36& 16.27 \\
	&      & 	    &	   & 	    &	    &		 &	&  	   & 48-58 & 17.36-21.04& 19.20 \\
\hline \\
Ch3E5  &  0-8  &  2.06-3.68 &  2.87&  Ch4E5  &  0-8  &  2.06-3.68 &  2.87&   Ch5E3  &  0-8  &  1.94-3.35 &  2.65 \\
       &  9-11 &  3.68-4.90 &  4.29& 	     &  9-11 &  3.68-4.90 &  4.29&  	    &  9-11 &  3.35-4.41 &  3.88 \\
       & 12-13 &  4.90-5.71 &  5.31& 	     & 12-13 &  4.90-5.71 &  5.31&  	    & 12-13 &  4.41-5.12 &  4.77 \\
       & 14-15 &  5.71-6.53 &  6.12& 	     & 14-15 &  5.71-6.53 &  6.12&  	    & 14-15 &  5.12-5.82 &  5.47 \\
       & 16-19 &  6.53-8.17 &  7.35& 	     & 16-19 &  6.53-8.17 &  7.35&  	    & 16-19 &  5.82-7.25 &  6.54 \\
       & 20-23 &  8.17-9.81 &  8.99& 	     & 20-23 &  8.17-9.81 &  8.99&  	    & 20-23 &  7.25-8.68 &  7.97 \\
       & 24-29 &  9.81-12.28& 11.05& 	     & 24-27 &  9.81-11.45& 10.63&  	    & 24-27 &  8.68-10.83&  9.40 \\
       & 30-35 & 12.28-14.76& 13.52& 	     & 28-35 & 11.45-14.76& 13.11&  	    & 28-35 & 10.83-12.99& 11.55 \\
       & 36-41 & 14.76-17.26& 16.01& 	     & 36-39 & 14.76-16.43& 15.60&  	    & 36-41 & 12.99-15.17& 14.08 \\
       & 42-47 & 17.26-19.78& 18.52& 	     & 40-45 & 16.43-18.94& 17.69&  	    & 42-47 & 15.17-17.36& 16.27 \\
       & 48-58 & 19.78-24.00& 21.89& 	     & 46-57 & 18.94-24.00& 21.47&  	    & 48-58 & 17.36-21.04& 19.20 \\
\hline \\
Ch6E3  &  0-8  &  1.94-3.35 &  2.65&  Ch6E4  &  0-8  &  2.13-3.79 &  2.96&   Ch6E5  &  0-8  &  2.06-3.68 &  2.87 \\
       &  9-11 &  3.35-4.41 &  3.88& 	     &  9-11 &  3.79-5.04 &  4.42&  	    &  9-11 &  3.68-4.90 &  4.29 \\
       & 12-13 &  4.41-5.12 &  4.77& 	     & 12-13 &  5.04-5.88 &  5.46&  	    & 12-13 &  4.90-5.71 &  5.31 \\
       & 14-15 &  5.12-5.82 &  5.47& 	     & 14-15 &  5.88-6.72 &  6.30&  	    & 14-15 &  5.71-6.53 &  6.12 \\
       & 16-19 &  5.82-7.25 &  6.54& 	     & 16-19 &  6.72-8.40 &  7.56&  	    & 16-19 &  6.53-8.17 &  7.35 \\
       & 20-23 &  7.25-8.68 &  7.97& 	     & 20-23 &  8.40-10.09&  9.25&  	    & 20-23 &  8.17-9.81 &  8.99 \\
       & 24-29 &  8.68-10.83&  9.76& 	     & 24-29 & 10.09-12.63& 11.36&  	    & 24-29 &  9.81-12.28& 11.05 \\
       & 30-35 & 10.83-12.99& 11.91& 	     & 30-35 & 12.63-15.19& 13.91&  	    & 30-35 & 12.28-14.76& 13.52 \\
       & 36-39 & 12.99-14.44& 13.72& 	     & 36-39 & 15.19-16.90& 16.05&  	    & 36-39 & 14.76-16.43& 15.60 \\
       & 40-46 & 14.44-17.00& 15.72& 	     & 40-46 & 16.90-19.92& 18.41&  	    & 40-46 & 16.43-19.36& 17.90 \\
       & 47-58 & 17.00-21.04& 19.02& 	     & 47-58 & 19.92-24.70& 22.31&  	    & 47-58 & 19.36-24.00& 21.68 \\

\enddata\label{table2}

\tablenotetext{a}{: Ch* in ChID represents a definition of PCA energy bands for light curve extraction, 
and E* in ChID represents the PCA Gain Epoch of the corresponding observations, e.g., Ch3E5 means the third definition in Epoch 5.}

\end{deluxetable}

\end{document}